\documentclass[aps,prb,twocolumn]{revtex4-1} % onecolumn,preprint,showpacs
\usepackage{graphicx,amsmath,multirow}
\begin{document}

% Use the \preprint command to place your local institutional report number 
% on the title page in preprint mode.
% Multiple \preprint commands are allowed.
%\preprint{}

\title{Crystal truncation rods from miscut surfaces} %Title of paper

% repeat the \author .. \affiliation  etc. as needed
% \email, \thanks, \homepage, \altaffiliation all apply to the current author.
% Explanatory text should go in the []'s, 
% actual e-mail address or url should go in the {}'s for \email and \homepage.
% Please use the appropriate macro for the type of information

% \affiliation command applies to all authors since the last \affiliation command. 
% The \affiliation command should follow the other information.

\author{Trevor A Petach}
\affiliation{Department of Physics, Stanford University, Palo Alto, CA 94305, USA}
\author{Apurva Mehta}
\affiliation{SSRL, SLAC National Accelerator Laboratory, Menlo Park, CA 94205, USA}
\author{Michael F Toney}
\affiliation{SSRL, SLAC National Accelerator Laboratory, Menlo Park, CA 94205, USA}
\author{David Goldhaber-Gordon}
\affiliation{Department of Physics, Stanford University, Palo Alto, CA 94305, USA}

%\email[]{Your e-mail address}
%\homepage[]{Your web page}
%\thanks{}
%\altaffiliation{}

% Collaboration name, if desired (requires use of superscriptaddress option in \documentclass). 
% \noaffiliation is required (may also be used with the \author command).
%\collaboration{}
%\noaffiliation

\date{\today}

\begin{abstract}
Crystal truncation rods are used to study surface and interface structure. Since real surfaces are always somewhat miscut from a low index plane, it is important to study the effect of miscut on crystal truncation rods. We develop a model that describes the truncation rod scattering from miscut surfaces that have steps and terraces. We show that non-uniform terrace widths and jagged step edges are both forms of roughness that decrease the intensity of the rods. Non-uniform terrace widths also result in a broad peak that overlaps the rods. We use our model to characterize the terrace width distribution and step edge jaggedness on three SrTiO$_3$ (001) samples, showing excellent agreement between the model and the data, confirmed by atomic force micrographs of the surface morphology. We expect our description of terrace roughness will apply to many surfaces, even those without obvious terracing.
\end{abstract}

\pacs{61.05.cc, 61.05.cf, 61.05.cp, 68.35.Ct}% insert suggested PACS numbers in braces on next line

\maketitle %\maketitle must follow title, authors, abstract and \pacs

% Body of paper goes here. Use proper sectioning commands. 
% References should be done using the \cite, \ref, and \label commands
%

%%%%%
\section{Introduction}
%\label{}
%\subsection{Subsection one}
Surface x-ray diffraction is a critical tool for understanding surface structure on an atomic scale. \cite{Robinson1992} One useful surface diffraction technique is analysis of crystal truncation rods, which are streaks of scattering extending away from the Bragg peaks parallel to the surface normal. Crystal truncation rods have been used to solve surface reconstructions, \cite{Barbier2000,Vonk2005} locate adatoms, \cite{Feidenhansl1990} study self-assembled monolayers, \cite{Fenter1994} and understand buried interfaces, \cite{Willmott2007,Yacoby2013} solid-liquid interfaces, \cite{Toney1994,Liu2016} and solid-gas interfaces. \cite{Eng2000} Thus, they are a critical tool for understanding surface structure determination.

There are two approaches to simulating the truncation rod intensity. The first ``continuum'' approach presumes that a crystal can be described as an infinite lattice multiplied by a shape function, which is unity in the bulk and zero outside the crystal. \cite{Andrews1985} The Fourier transform of the shape function determines the shape of the truncation rod. The second ``atomistic'' approach is to add up the scattering from every atom in the crystal, with an appropriate phase factor that depends on the position of the atom. \cite{Robinson1986} The square of the magnitude of the sum is proportional to the truncation rod intensity.

Roughness can be included in both models. In the continuum approach, roughness is captured by a broadening of the shape function. \cite{Sinha1988} In the atomistic approach, roughness is modeled as a series of partially occupied layers near the surface. In the best-known formulation, called $\beta$-roughness, the occupancy of each layer above the bulk is a constant fraction, $\beta$, of the layer below. \cite{Robinson1986} In both approaches, roughness reduces the intensity of the truncation rod, with the largest effect at the anti-Bragg points. Other models have also been developed. For example, co-existence of two-dimensional and three-dimensional growth modes in a thin film results in a more complex roughness factor. \cite{Shi2016} These approaches generally work well when the crystal surface is parallel to a low-index plane.

However, no real surface is parallel to a low-index plane. Even if the surface is locally parallel to the plane, steps, often one unit cell tall, divide the surface into terraces whose lateral spacing depends on the miscut angle. Provided the coherence length exceeds the terrace width, a separate ``sub-rod'' will extend from each Bragg point, as shown in Fig.~\ref{fig:Overview}(a). Measuring the truncation rods from such a surface with a point detector is challenging because each sub-rod must be measured separately, and small misalignments or imperfections in the diffractometer require frequent alignment scans to find the sub-rods. Thus, experimentalists frequently use samples with small miscuts ($<0.05^{\circ}$) and align the miscut with the low-resolution direction of the beam or diffractometer to avoid the need to track the individual sub-rods.\cite{Vonk2005, Dale2006} Recently, area detectors have made the task of measuring truncation rods from miscut samples much easier since the detector usually intercepts several sub-rods simultaneously and small misalignments have minimal impact on the data collection.\cite{Schleputz2005} In light of this much easier data collection, it is necessary to develop a theory of truncation rods from miscut samples so that a wider variety of samples can be studied. 

In this paper, we develop a model for the crystal truncation rods from miscut surfaces with terraces and steps and show that it agrees well with data collected from miscut SrTiO$_3$ (001) surfaces, whose morphology is separately characterized by atomic force microscopy.

%%%%%
\section{Model}

It is well known that scattering from a surface encodes information about the height-height correlation function.  \cite{Lu1982,Sinha1988} Several models have been developed to describe diffraction from vicinal, stepped surfaces that include a variety of terrace width distributions and step edge roughnesses. \cite{Presicci1984,Pukite1985,Croset1998,Wollschlager2007} Experimentally, the details of the step distribution can be elucidated using these models. For example, step edge repulsion and phase separation were observed in miscut silicon, \cite{Held1995,Held1995b} and anisotropy in a roughening transition was observed in Ag (110). \cite{Pflanz1995}

These models for step distributions generally don't describe the intensity along the entire truncation rod, but rather only the in-plane shape. However, it has been shown in a simple model with two terrace widths that unequal terrace widths reduce the truncation rod intensity, especially at the anti-Bragg points. \cite{Munkholm1999} A numerical calculation of truncation rod intensities from an atomic force microscope image of a terraced, miscut surface shows a similar effect. \cite{Dale2006}

\begin{figure}[htp!]
\includegraphics[width=3.375in]{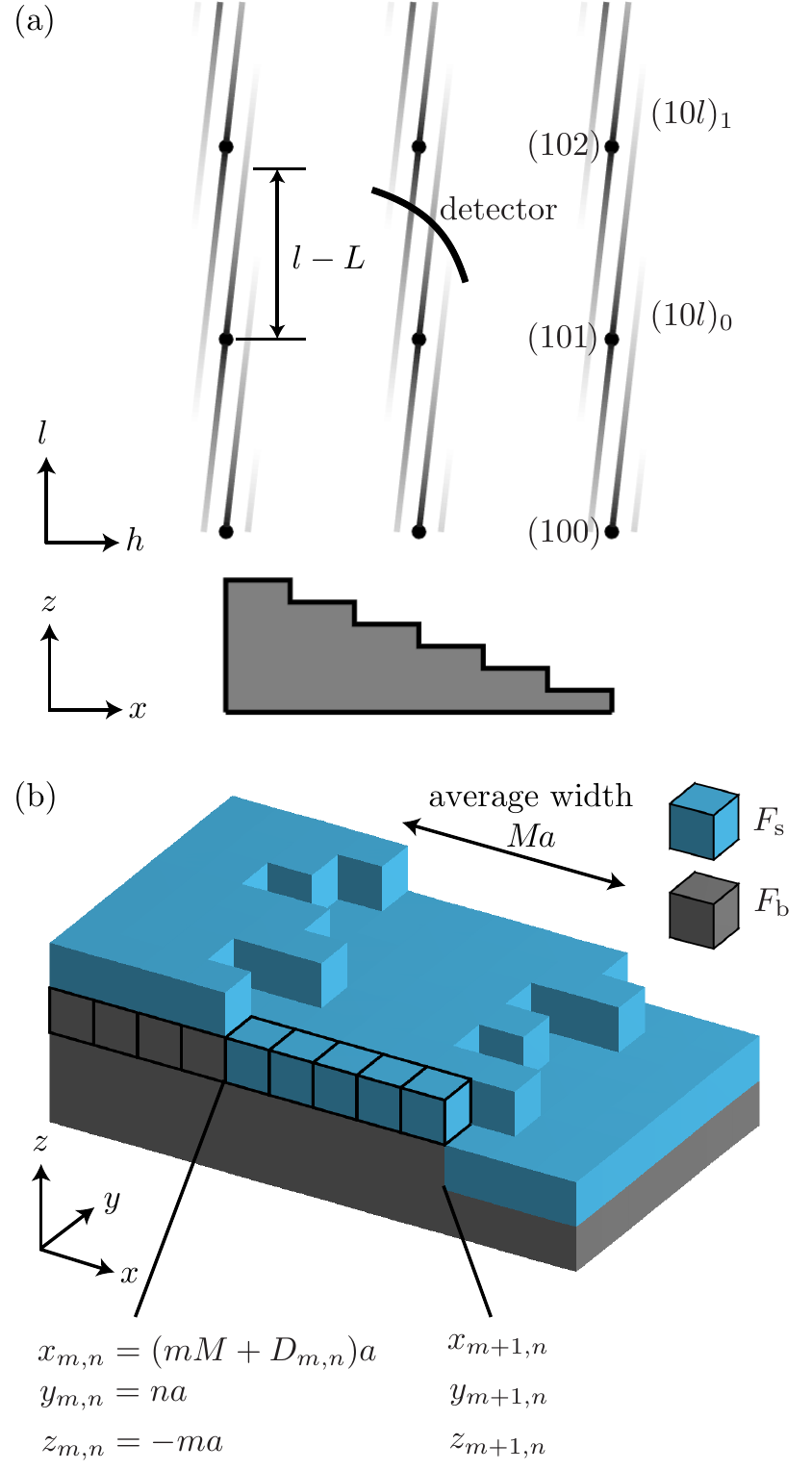}
\caption{Schematic of the truncation rods and the sample surface. (a) Miscut results in splitting of the truncation rod, with separate sub-rods extending away from each Bragg point in a direction perpendicular to the surface. $L$ is an integer labeling the out of plane index of the Bragg points. We label each sub-rod with a subscript indicating the Bragg point from which it emanates. It is possible to intercept all sub-rods from a given rod using an area detector. (b) Terraces are one bulk unit cell tall with an average width $Ma$. The deviation from the zero-roughness position of the step edge on the $m$th terrace, $n$ unit cells along the step is $D_{m,n}a$. $F_{\mathrm{s}}$ and $F_{\mathrm{b}}$ are the the structure factors of the surface unit cell and bulk unit cell, respectively.}
\label{fig:Overview}
\end{figure}

Building on these principles, we develop an atomistic model of a miscut surface, presuming a cubic crystal comprised of bulk unit cells (with structure factor $F_{\mathrm{b}}$) covered by a single layer of a different surface unit cell (with structure factor $F_{\mathrm{s}}$). We presume that all step edges are one bulk unit cell tall and that the average terrace width is $M$ unit cells, so the miscut angle is $\arctan(1/M)$. However, we do not presume that the terraces edges are straight, or that each terrace has the same width. As shown in Fig.~\ref{fig:Overview}(b), the position of the step edge at the end of the $m$th terrace, $n$ unit cells along the step, is $((m+1)M + D_{m,n})a$, where $D_{m,n}$ is the deviation of the position from the ideal value and $a$ is the lattice constant. For an ideal surface with straight-edged, uniformly spaced terraces, all $D_{m,n} = 0$. 

To begin, we add the structure factors from a single row of unit cells, outlined in black in Fig.~\ref{fig:Overview}(b), to find

\begin{multline}
F_{m,n} = \left( \sum_{j=-\infty}^{x_{m,n}/a-1} F_{\mathrm{b}} e^{i q_x a j} + \sum_{j=x_{m,n}/a}^{x_{m+1,n}/a-1} F_{\mathrm{s}} e^{i q_x a j} \right) \\  \times \vphantom{\sum_{j=-\infty}^{x_{m,n}-1}} e^{iq_y a n} e^{-iq_z a m}.
\end{multline}

%\begin{equation}
%F_{row,m,n} = -\frac{F_{\mathrm{b}}}{1-e^{iq_x a}} + \frac{F_{\mathrm{s}}\left(1-e^{iq_x a(M+D_{m+1,n}-D_{m,n})}\right)}{1-e^{iq_x a}}
%\end{equation}

%\begin{equation}
%F_{patch} = \sum_{n=-\infty}^{\infty}  \sum_{m=-\infty}^{\infty} F_{row,m,n}e^{iq_x a(mM+D_{m,n})} e^{iq_y a n} e^{-iq_z a m} e^{-m^2/S^2} e^{-n^2/T^2}
%\end{equation}

If the beam were perfectly coherent, the scattered amplitudes from the entire illuminated surface would add coherently. To account for partial coherence, we add scattered amplitudes from a local region or ``patch,'' weighting amplitudes farther from the center of the patch less than those near the center:
\begin{multline}
F_{\mathrm{patch}} = \\
\sum_{n=-\infty}^{\infty}  \sum_{m=-\infty}^{\infty} F_{m,n} e^{-m^2 M^2a^2/\xi_x^2} e^{-n^2a^2/\xi_y^2},
\end{multline}
where $\xi_x$ and $\xi_y$ are the coherence lengths in the $x$ and $y$ directions. Presuming that any correlation in the deviations $D_{m,n}$ decays on a shorter scale than either coherence length, the scattered intensity is proportional to

\begin{equation}
I = \frac{Aa^2}{2\pi^3 \xi_x \xi_y} \langle F_{\mathrm{patch}} F_{\mathrm{patch}}^* \rangle,
\label{eq:ave}
\end{equation}
where $A$ is the illuminated area, and the brackets denote the spatial average over the whole sample. Expanding this expression, we find

\begin{multline}
I = \frac{Aa^2}{2\pi^3 \xi_x \xi_y} \left\langle \sum_{m,n,m',n'} F_{m,n} F^*_{m',n'} \right. \\
\times \left. e^{-(m^2+m'^2)M^2a^2/\xi_x^2} e^{-(n^2+n'^2)a^2/\xi_y^2} \vphantom{\sum_{m,n,m',n'}} \right\rangle.
\end{multline}
Defining
\begin{equation}
\tilde{F}_{m,n} \equiv \frac{\left(F_{\mathrm{s}} - F_{\mathrm{b}}\right) e^{i q_x a D_{m,n}} - F_{\mathrm{s}} e^{iq_x a(M+D_{m+1,n})}}{1-e^{iq_x a}}
\end{equation}
and
\begin{equation}
\theta \equiv q_x a M - q_z a,
\end{equation}
the expression for intensity becomes
\begin{multline}
I = \frac{Aa^2}{2\pi^3 \xi_x \xi_y} \sum_{m,n,m',n'} \left\langle \tilde{F}_{m,n} \tilde{F}^*_{m',n'} \right\rangle e^{i\theta(m-m')}  \\
\times \vphantom{\sum_{m,n,m',n'}} e^{i q_y a (n-n')} e^{-(m^2+m'^2)M^2a^2/\xi_x^2} e^{-(n^2+n'^2)a^2/\xi_y^2},
\label{eq:sum}
\end{multline}

where

\begin{align}
\left\langle \tilde{F}_{m,n} \right. & \left. \tilde{F}^*_{m',n'} \right\rangle  =  \nonumber \\
 &\left|F_{\mathrm{s}}-F_{\mathrm{b}}\right|^2 \left\langle e^{iq_x a(D_{m,n}  - D_{m',n'})} \right\rangle \nonumber \\
- &(F_{\mathrm{s}}-F_{\mathrm{b}})F_{\mathrm{s}}^* e^{-i q_x a M} \left\langle e^{i q_x a(D_{m,n}-D_{m'+1,n'})}  \right\rangle \nonumber \\
- &(F_{\mathrm{s}}-F_{\mathrm{b}})^*F_{\mathrm{s}} e^{i q_x a M} \left\langle e^{i q_x a (D_{m+1,n} - D_{m',n'})} \right\rangle \nonumber \\
+ &\left|F_{\mathrm{s}}\right|^2 \left\langle  e^{i q_x a(D_{m+1,n}-D_{m'+1,n})} \right\rangle.
\end{align}

In order to simplify the calculation, we assume that the deviations $D_{m,n}$ have zero mean and a Gaussian distribution. Then, we can use the Baker-Hausdorff Theorem \cite{Nielsen2011} to calculate the spatial average,

\begin{equation}
\left \langle e^{i q_x a (D_{m,n}-D_{m',n'})} \right\rangle = e^{-(q_x a)^2 \langle (D_{m,n}-D_{m',n'})^2 \rangle/2}.
\label{eq:baker}
\end{equation}

%\begin{equation}
%\langle (D_{m,n})^2 \rangle = \langle (D_{m',n'})^2 \rangle = \sigma_{\mathrm{s}}^2 + \sigma_{\mathrm{w}}^2
%\end{equation}

\begin{figure}[htp!]
\includegraphics[width=2.92in]{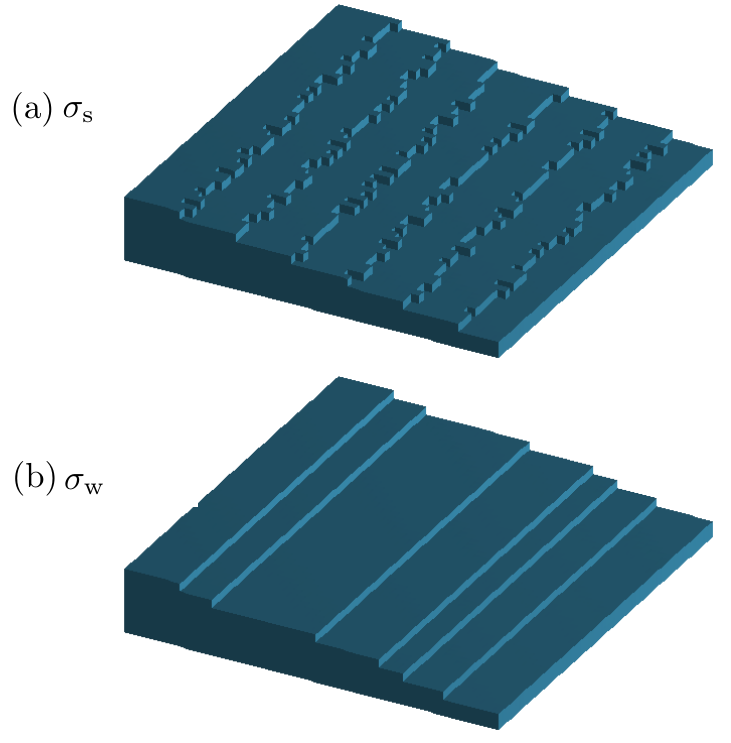}
\caption{Roughness on a terraced surface can arise from (a) step edge jaggedness and/or (b) width variation from terrace to terrace.}
\label{fig:sigma}
\end{figure}

To proceed, we must calculate the average difference between the step positions at different locations, $\langle (D_{m,n}-D_{m',n'})^2 \rangle$. For most surfaces, this quantity is a complicated function of $m-m'$ and $n-n'$ that depends on the details of the step distribution on that particular surface. In order to proceed, we use a simple step distribution. As shown in Fig.~\ref{fig:sigma}, we presume that there are only two non-idealities in the step edges. First, any single step edge is jagged, with standard deviation from the average position $\sigma_{\mathrm{s}}$ (``s'' for ``step'') and no correlation in the jaggedness along the step. Second, the terrace width changes from terrace to terrace, with standard deviation from the average width $\sigma_{\mathrm{w}}$ (``w'' for ``width'') and no correlation between widths on subsequent terraces. These two types of roughness result in the average correlation function

\begin{equation}
\frac{\langle (D_{m,n}-D_{m',n'})^2 \rangle}{2} = \begin{cases} 
      0 & n=n', m=m' \\
      \dfrac{\sigma_{\mathrm{s}}^2}{a^2} & n \neq n', m=m' \\
      \dfrac{\sigma_{\mathrm{s}}^2 + \sigma_{\mathrm{w}}^2}{a^2} & m \neq m' \\
\end{cases}
\label{eq:corr}
\end{equation}

\begin{figure}[htp!]
\centering
\includegraphics[width=3.375in]{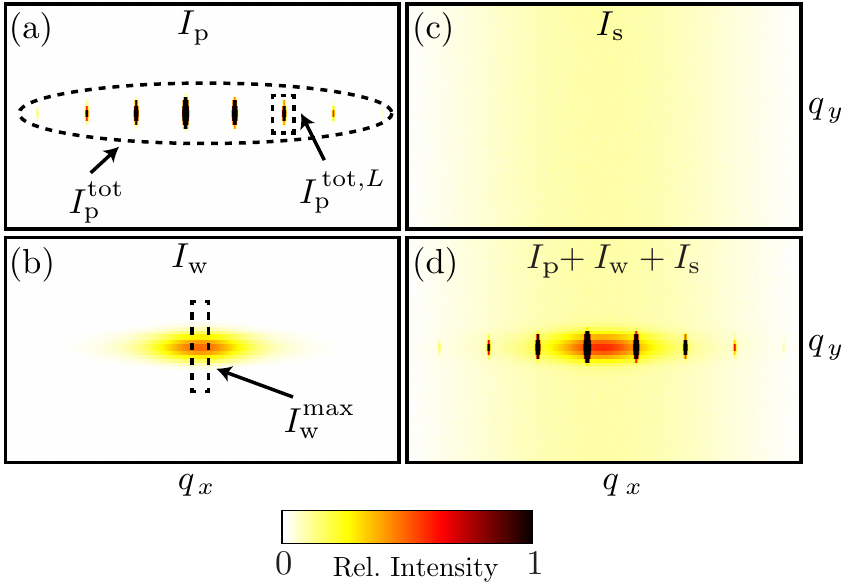}
\caption{The three components of the scattered intensity from a miscut surface at fixed $q_z$. (a) $I_{\mathrm{p}}$ -- sharp peaks given by Eq.~(\ref{eq:pks}). $I_{\mathrm{p}}^{\mathrm{tot},L}$ is the intensity in a single peak integrated over $q_x$ and $q_y$. $I_{\mathrm{p}}^{\mathrm{tot}}$ is the sum of the integrated intensity in all of the peaks. (b) $I_{\mathrm{w}}$ -- a broad peak that arises from variations in terrace width from terrace to terrace given by Eq.~(\ref{eq:w}). We define $I_{\mathrm{w}}^{\rm{max}}$ as the sum of the column containing the highest intensity pixel. (c) $I_{\mathrm{s}}$ -- a diffuse background that arises from jagged step edges given by Eq.~(\ref{eq:s}). The intensity is usually too low to observe experimentally. (d) The total scattered intensity.}
\label{fig:schematic}
\end{figure}
 
Many surfaces will have much more complicated forms of this correlation function. However, this simple form captures the main features of many terraced surfaces. Evaluating the sum in Eq.~(\ref{eq:sum}) using Eq.~(\ref{eq:baker}) and Eq.~(\ref{eq:corr}), we find three distinct components of the total intensity
\begin{equation}
I = I_{\mathrm{p}} + I_{\mathrm{w}} + I_{\mathrm{s}},
\label{eq:I}
\end{equation}
where $I_{\mathrm{p}}$, $I_{\mathrm{w}}$, and $I_{\mathrm{s}}$ are all functions of $q_x$, $q_y$, and $q_z$, $F_{\mathrm{b}}$ and $F_{\mathrm{s}}$, and the two roughness parameters $\sigma_{\mathrm{s}}$ and $\sigma_{\mathrm{w}}$. These functions are shown in Fig.~\ref{fig:schematic} for fixed $q_z$. Eq.~(\ref{eq:I}) is a significant result. It states that the scattering from a miscut surface can be divided into three distinct parts arising from 1) the splitting of the truncation rod due to the miscut, 2) a broad peak due to variable terrace widths, and 3) a diffuse background from jagged step edges. We now discuss these parts in more detail.

\subsection{$I_{\mathrm{p}}$ -- Sharp Peaks from Sub-Rods}

Most of the scattering in our model comes in a series of sharp peaks, given by

\begin{multline}
I_{\mathrm{p}} = \vphantom{\frac{1}{1}} \frac{Aa^2}{2\pi^3 \xi_x \xi_y} H_0(\theta,\xi_x/Ma) H_0(q_y a,\xi_y/a) \\ 
\times \left| (F_{\mathrm{s}}-F_{\mathrm{b}}) - F_{\mathrm{s}} e^{i q_x a M} \right|^2 \frac{e^{-q_x^2 (\sigma_{\mathrm{s}}^2 + \sigma_{\mathrm{w}}^2)}}{4\sin^2(q_x a/2)},
\label{eq:pks}
\end{multline}
(with ``p'' for ``peaks''), where

\begin{equation}
H_0(x,b) = \sum_{j=-\infty}^{\infty} \pi b^2 e^{-\frac{1}{2} b^2 (x-2\pi j)^2}.
\end{equation}
Eq.~\eqref{eq:pks} is plotted in Fig.~\ref{fig:schematic}(a) for fixed $q_z$. The function $H_0(x,b)$ is a periodic series of peaks spaced by $2\pi$ in $x$. Thus, the product of the two $H_0$ functions in Eq.~(\ref{eq:pks}) restricts scattering to a series of rods in reciprocal space. With no miscut, there would be a single rod for each integer value of $h$ and $k$, and the rods would be labeled $(00l)$, $(10l)$, and so on, in the usual representation. As shown in Fig.~\ref{fig:Overview}(a), miscut splits these rods into separate sub-rods each emanating from a single Bragg point, which we call the ``primary'' Bragg point for that sub-rod. Thus, we label each sub-rod by the usual notation plus a subscript noting the $l$ value of the primary Bragg point, so the $(10l)$ rod now splits into several sub-rods, labeled $(10l)_0$, $(10l)_1$, and so on. An area detector often intercepts several of these sub-rods simultaneously, displaying a series of sharp peaks. These peaks will be spaced by $2\pi/Ma$ in $q_x$. The larger the miscut, the larger the spacing between peaks.

We have assumed that the coherence length is significantly longer than the average terrace width, so that the peaks from the sub-rods are well defined. If the coherence length is shorter than or comparable to the average terrace width, then the width of each sub-rod will be broad enough that the individual sub-rods will be indistinguishable, and the scattering will appear like a single rod connecting Bragg peaks in the out-of-plane direction. In that case, the analysis is more complicated.

The second line of Eq.~(\ref{eq:pks}) modulates the intensity of the sub-rods in two ways. First, there is the standard interference between the surface and the bulk, in which the bulk dominates near the primary Bragg point (and whenever $l$ is an integer), whereas they contribute equally whenever $l$ is a half-integer. Second, there is a roughness factor which reduces the intensity of the sub-rod away from the primary Bragg point. Larger terrace width variation, $\sigma_{\mathrm{w}}$, and larger step edge jaggedness, $\sigma_{\mathrm{s}}$, both result in a faster reduction of intensity moving away from the primary Bragg point.

Experimentally, in a truncation rod measurement, $I_{\mathrm{p}}$ will be observed as a series of peaks on an area detector, one from each sub-rod. One useful way to treat such data is to subtract a background from each peak and then add the total intensity in all of the peaks. To calculate the total integrated intensity observed in this case, we need to integrate over $q_x$ and $q_y$. We note that
\begin{equation}
\int_{-\infty}^{\infty} \pi b^2 e^{-\frac{1}{2} b^2 x^2} dx = \pi \sqrt{2\pi} b.
\end{equation}

Even though $F_{\mathrm{s}}$ and $F_{\mathrm{b}}$ are functions of $q_x$ and $q_y$, we treat them as constants during the integration since only a few peaks contribute to the integral and $F_{\mathrm{s}}$ and $F_{\mathrm{b}}$ vary only slightly from peak to peak (for $M >> 1$). Since $H_0(\theta,\xi_x/Ma)$ is peaked at $\theta = 2 \pi L$, where $L$ is an integer, we approximate it as a series of $\delta$-functions and make the substitution $q_x = (q_z a - 2 \pi L)/Ma$, which allows us to write the result as a sum:

\begin{multline}
I_{\mathrm{p}}^{\mathrm{tot}} = \frac{A}{M^2a^2} \left| (F_{\mathrm{s}}-F_{\mathrm{b}}) - F_{\mathrm{s}} e^{i q_z a} \right|^2 \\ 
\times \sum_L \frac{ e^{-(q_z a - 2\pi L)^2(\sigma_{\mathrm{tot}}/Ma)^2 } }{4\sin^2\big((q_z a - 2\pi L)/2M\big)},
\label{eq:Ipkssum}
\end{multline}
where 

\begin{equation}
\sigma_{\mathrm{tot}} \equiv \sqrt{\sigma_{\mathrm{s}}^2 + \sigma_{\mathrm{w}}^2}
\end{equation}
is the total ``terrace roughness.'' For $M>>1$, only terms near $L = q_z a/2\pi$ contribute to the sum, so the argument of $\sin x$ in the denominator is small. Expanding and rearranging, we find

\begin{multline}
I_{\mathrm{p}}^{\mathrm{tot}} = \frac{A}{a^2} \left| F_{\mathrm{s}} + F_{\mathrm{b}} \frac{e^{-i q_z a}}{1-e^{-i q_z a}} \right|^2 \\
\times \sum_L 4 \frac{\sin^2(q_z a/2)}{(q_z a - 2\pi L)^2} e^{-(q_z a - 2\pi L)^2\tilde{\sigma}_{\mathrm{tot}}^2},
\label{eq:Ipkstot}
\end{multline}
where $\tilde{\sigma}_{\mathrm{tot}} \equiv \sigma_{\mathrm{tot}}/Ma$ is the total roughness as a fraction of the terrace length, and $\tilde{\sigma}_{\mathrm{w}}$ and $\tilde{\sigma}_{\mathrm{s}}$ are similarly defined. Each term in the sum in Eq.~(\ref{eq:Ipkstot}) is the integrated intensity in the sharp peak from the $L$th sub-rod, which we denote $I_{\mathrm{p}}^{\mathrm{tot},L}$.

In order to isolate the effect of roughness, we define
\begin{equation}
I_0 \equiv \frac{A}{a^2} \left| F_{\mathrm{s}} + F_{\mathrm{b}} \frac{e^{-i q_z a}}{1-e^{-i q_z a}} \right|^2
\end{equation}
and a roughness factor

\begin{align}
c_{\mathrm{p}} & \equiv \sum_L 4 \frac{\sin^2(q_z a/2)}{(q_z a - 2\pi L)^2} e^{-(q_z a - 2\pi L)^2\tilde{\sigma}_{\mathrm{tot}}^2 } \\
& \approx 1 - \left( \frac{4}{\sqrt{\pi}} \tilde{\sigma}_{\mathrm{tot}} + \mathcal{O}(\tilde{\sigma}_{\mathrm{tot}}^3) \right) \sin^2\Big(\frac{q_z a}{2}\Big),
\label{eq:approx}
\end{align}
where the first order approximation in Eq.~(\ref{eq:approx}) is valid for $\tilde{\sigma}_{\mathrm{tot}} \lesssim 0.3$ (which limits the error to 3\%).\cite{SM} The expression for $I_{\mathrm{p}}^{\mathrm{tot}}$ can then be written

\begin{equation}
I_{\mathrm{p}}^{\mathrm{tot}} = c_{\mathrm{p}} I_0,
\end{equation}
where $I_0$ is the intensity had there been no miscut, and $c_{\mathrm{p}}$ is a factor that depends only on $q_z$ and $\sigma_{\mathrm{tot}}$ and not on $F_{\mathrm{b}}$ or $F_{\mathrm{s}}$. It is unity when the total roughness is zero and less than one otherwise. (We discuss the roughness factor in detail in Sec.~\ref{sec:Discussion}.) Thus, the total integrated intensity in all of the peaks is proportional to the intensity had there been no miscut and a roughness factor which depends only on $q_z$ and $\sigma_{\mathrm{tot}}$.

\subsection{$I_{\mathrm{w}}$ -- Broad Peak}

If the terraces have non-uniform widths, regardless of whether the step edges are straight or jagged, then a broad peak develops underneath the sharp peaks, described by 

\begin{multline}
I_{\mathrm{w}} =  \frac{A a \sqrt{\pi/2}}{2\pi^3 \xi_y M} H_0(q_y a,\xi_y/a) \left| (F_{\mathrm{s}}-F_{\mathrm{b}}) - F_{\mathrm{s}} e^{i q_z a} \right|^2 \\
\times \frac{ e^{-q_x^2 \sigma_{\mathrm{s}}^2} - e^{-q_x^2 (\sigma_{\mathrm{s}}^2 + \sigma_{\mathrm{w}}^2)} }{4\sin^2(q_x a/2)}.
\label{eq:w}
\end{multline}

Eq.~(\ref{eq:w}) is plotted in Fig.~\ref{fig:schematic}(b) for fixed $q_z$. The width in $q_y$ is inversely proportional to the coherence length. However, the width in $q_x$ is much wider than the inverse of the coherence length and depends on the terrace width nonuniformity. When the width variation $\sigma_{\mathrm{w}}$ is small, the peak is weak and broad in $q_x$. As $\sigma_{\mathrm{w}}$ increases, the peak becomes narrower and stronger in such a way that the integrated intensity increases. Step edge jaggedness ($\sigma_{\mathrm{s}}$) reduces the peak width in $q_x$ but does not change the maximum intensity. In any crystal with uniform terrace widths (but where the step edges may or may not be jagged), $\sigma_{\mathrm{w}}$ is zero and $I_{\mathrm{w}}$ is zero.

Experimentally, the scattering from a sample with nonuniform terrace widths is a series of sharp peaks from the sub-rods with a broad peak underneath. One approach to analyzing such data is to integrate the total intensity in all of these peaks. To calculate the total integrated intensity observed in this case, we need to integrate $I_{\mathrm{w}}$ over $q_x$ and $q_y$ and add the result to $I_{\mathrm{p}}^{\mathrm{tot}}$. Again treating $F_{\mathrm{s}}$ and $F_{\mathrm{b}}$ as constants since they vary only slightly over the extent of the broad peak in $q_x$ and $q_y$, we find that the integrated intensity in the broad peak is

\begin{multline}
I_{\mathrm{w}}^{\mathrm{tot}} = \frac{A}{a^2} \left| F_{\mathrm{s}} + F_{\mathrm{b}} \frac{e^{-i q_z a}}{1-e^{-i q_z a}} \right|^2 \\ 
\times \sin^2(q_z a/2)\frac{4}{\sqrt{\pi}} \left( \tilde{\sigma}_{\mathrm{tot}} - \tilde{\sigma}_{\mathrm{s}} \right).
\label{eq:Iw}
\end{multline}

Defining

\begin{equation}
c_{\mathrm{w}} \equiv \sin^2(q_z a/2)\frac{4}{\sqrt{\pi}} \left( \tilde{\sigma}_{\mathrm{tot}} - \tilde{\sigma}_{\mathrm{s}} \right),
\end{equation}
the expression for $I_{\mathrm{w}}^{\mathrm{tot}}$ can be written

\begin{equation}
I_{\mathrm{w}}^{\mathrm{tot}} = c_{\mathrm{w}} I_0,
\end{equation}

where $c_{\mathrm{w}}$ is a factor that depends on $\sigma_{\mathrm{w}}$, $\sigma_{\mathrm{s}}$, and $q_z$, and not on $F_{\mathrm{b}}$ and $F_{\mathrm{s}}$. Thus, the integrated intensity in the broad peak is always proportional to the scattering had there been no miscut. Depending on how the detector images from an experiment are analyzed, this scattering may or may not need to be included during modeling. As we discuss in Sec.~\ref{sec:Discussion}, including it changes the shape of the rod in $q_z$.

\subsection{$I_{\mathrm{s}}$ -- Diffuse Background}

When the step edges are jagged, there is a diffuse background. It does not depend on $\sigma_{\mathrm{w}}$ and has the functional form

\begin{equation}
I_{\mathrm{s}} = \frac{A}{4\pi^2 M} \left| (F_{\mathrm{s}}-F_{\mathrm{b}}) - F_{\mathrm{s}} e^{i q_z a} \right|^2 \frac{ 1 - e^{-q_x^2 \sigma_{\mathrm{s}}^2} }{4\sin^2(q_x a/2)}.
\label{eq:s}
\end{equation}

Eq.~(\ref{eq:s}) is shown in Fig.~\ref{fig:schematic}(c) for constant $q_z$. This background is broad in all directions and cannot be easily measured experimentally. It would be subtracted in most reasonable background subtraction procedures.

%%%%%
\section{Experimental Results}
\label{sec:exp}

\begin{figure*}[htp!]
\includegraphics[width=\textwidth]{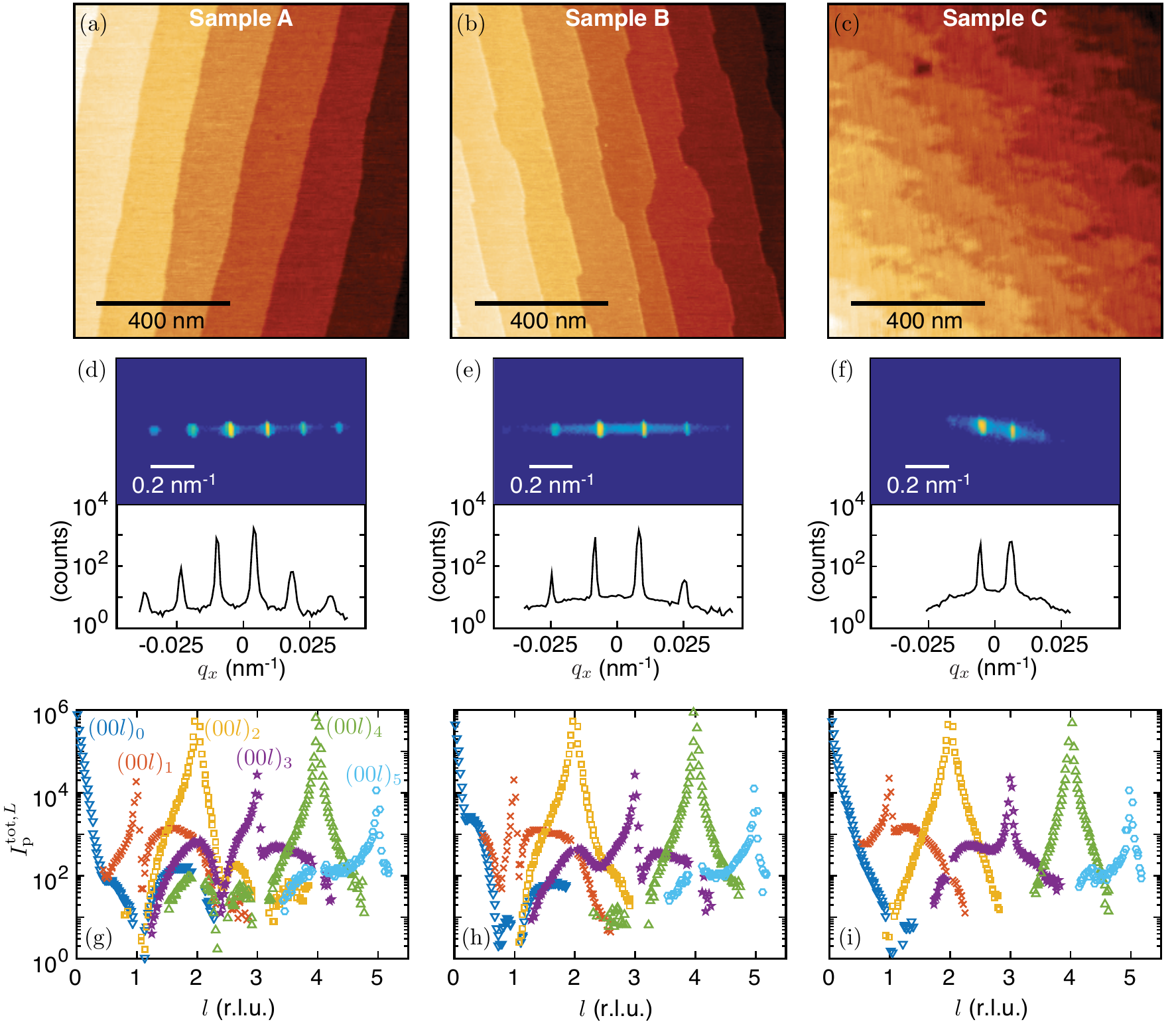}
\caption{Specular truncation rod from miscut SrTiO$_3$ (001). Atomic force micrographs of samples annealed at (a) 1025 C, (b) 950 C, and (c) no anneal, show different terrace morphologies. Each step is one unit cell (3.905 \AA) high. (d), (e), (f) Detector images where the center pixel is $l$ = 1.6, with column sums below. The scale bar on the image indicates the change in the length of $q$ across the image, whereas the $x$-axis labels on the plot indicate the change in $q_x$ only. (g), (h), (i) The integrated intensity of the sub-rods $L = 0$ to $L = 5$, plotted as a function of $l$. Each sub-rod is a different colored symbol. The Supplemental Material \cite{SM} contains a complete discussion of the correction factors applied to the data to obtain $I_{\mathrm{p}}^{\mathrm{tot},L}$.}
\label{fig:data}
\end{figure*}

To test our model, we prepared three SrTiO$_3$ (001) surfaces with different terrace morphologies that correspond to different values of $\sigma_{\mathrm{w}}$ and $\sigma_{\mathrm{s}}$. We call the samples A, B, and C. We etched the samples in 1:6 buffered oxide etch for 2 minutes to achieve TiO$_2$ termination. \cite{Kawasaki1993} We then annealed the samples differently: A at 1025 C in 1:10 O$_2$:Ar for 1 hr, B at 950 C in 1:10 O$_2$:Ar for 1 hr, C no anneal. These three annealing conditions resulted in three different surface morphologies, as shown by atomic force microscopy (AFM) in Fig.~\ref{fig:data}(a)-(c).

We measured the specular crystal truncation rod from the three samples at beamline 7-2 at SSRL in four circle mode with a double crystal Si (111) monochromator and a Rh-coated mirror to focus the beam to a spot approximately 100 x 500 $\mu$m FWHM. The energy was 15.5 keV. Scattered photons were collected on a Pilatus 100k area detector located approximately 1 m from the diffractometer center. Lorentz and illuminated area corrections were applied to all data. \cite{Busing1967,Vlieg1997,You1999} The Supplemental Material \cite{SM} contains a complete discussion of the corrections.

The scattering from the different surfaces agrees qualitatively with our theory. For all three samples, there are several sharp peaks and a single broad peak, as shown in Fig.~\ref{fig:data}(d)-(f). For Sample A, with the smoothest step edges and the most uniform widths, the sharp peaks fall off most slowly away from the center, and the broad peak is only faintly visible. The larger width variations in Sample B result in a stronger, narrower broad peak, and a faster falloff in intensity of the sharp peaks. For Sample C, the broad peak is similar to sample B, suggesting that the terrace width variance is similar, but the sharp peaks fall off more quickly, consistent with the more jagged step edges. The extra diffuse background from the jagged step edges, $I_{\mathrm{s}}$, is too weak to be visible. The scattering pattern is rotated for Sample C because the miscut direction is rotated relative to the crystal axes (see Supplemental Material \cite{SM} for further discussion of miscut rotation).

\begin{figure*}[htp!]
\includegraphics[width=\textwidth]{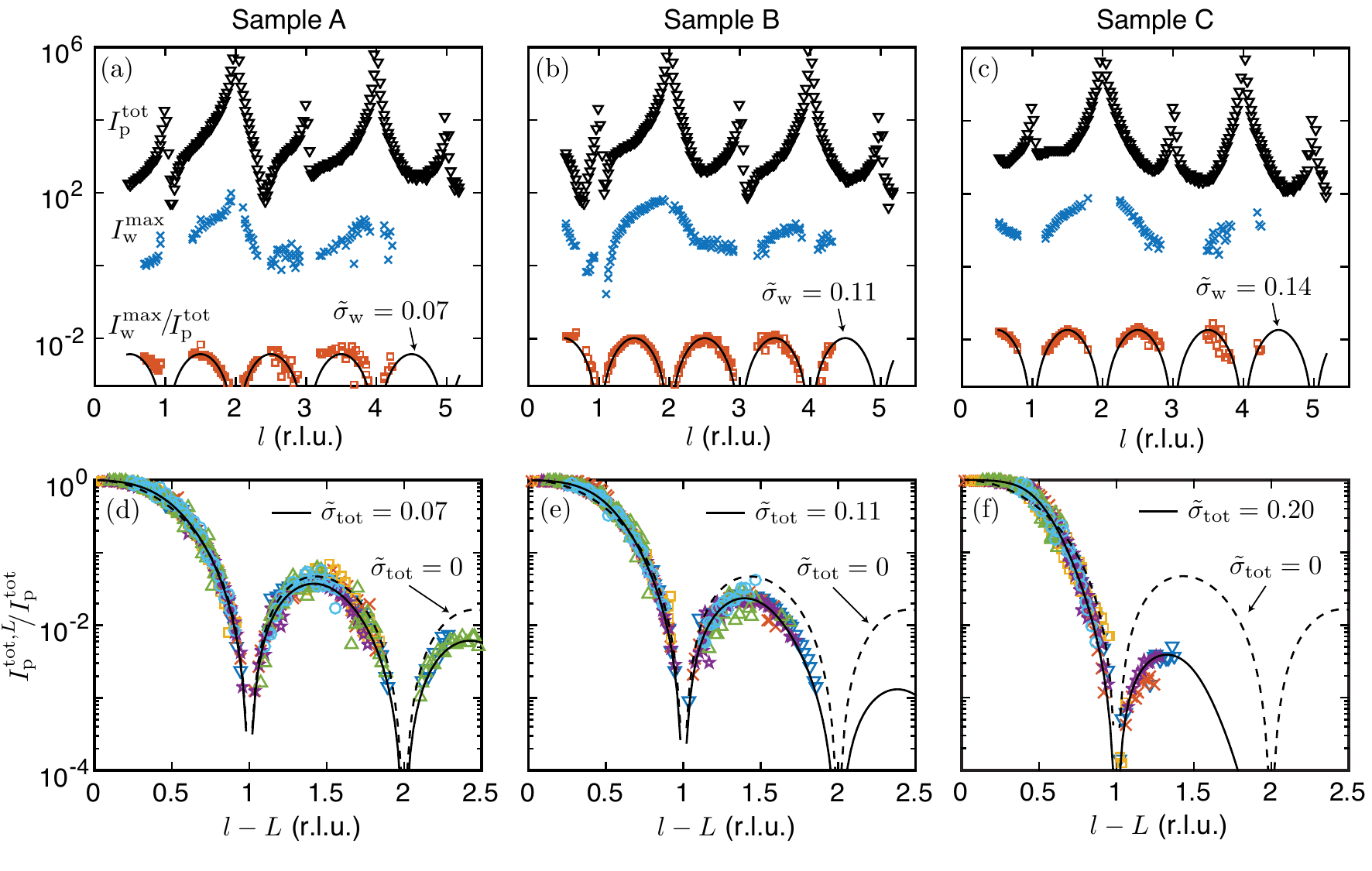}
\caption{Extracting roughness parameters. (a), (b), (c) The total intensity in the sharp peaks ($I_{\mathrm{p}}^{\mathrm{tot}}$, black triangles) and the height of the broad peak ($I_{\mathrm{w}}^{\rm{max}}$, blue crosses). The black lines are the best fit to the ratio of these intensities using Eq.~(\ref{eq:wratio}), where $\tilde{\sigma}_{\mathrm{w}}$ is the only fitting parameter. (d), (e), (f) The ratio of integrated intensity in a single sharp peak (plotted in Fig.~\ref{fig:data} (g)-(i)) to the total intensity in the sharp peaks, as a function distance from the primary Bragg peak. The black lines are the best fit to the ratio of these intensities using Eq.~(\ref{eq:pkratio}), where $\tilde{\sigma}_{\mathrm{tot}}$ is the only fitting parameter.}
\label{fig:analysis}
\end{figure*}

To make a quantitative estimate of $\sigma_{\mathrm{w}}$ and $\sigma_{\mathrm{s}}$, we compare both the integrated intensity of individual sharp peaks and the height of the broad peak to the total intensity in all the sharp peaks. 

Each sharp peak arises from the intersection of one sub-rod with the detector. Each sub-rod emanates from its primary Bragg point as an elliptic cylinder with its axis slightly tilted relative to the [00$l$] direction, where $l = q_za/2\pi$. For the specular rod, the primary Bragg points are located at (00$L$), where $L$ is an integer. We label the sub-rods using these integers. We plot the integrated intensities of the sharp peaks from sub-rods $L = 0$ to $L=5$ in Fig.~\ref{fig:data}(g)-(i). The integrated intensity in each sub-rod reaches a maximum at its primary Bragg point.

The detector occupies a region of the Ewald sphere, and thus in general intersects each sub-rod at a slightly different value of $l$, as shown in Fig.~\ref{fig:Overview}(a). However, for small miscut ($Ma >>$ thickness of surface unit cell), the intensity in the sub-rods varies slowly with $l$, and we approximate the intersection as occurring at the same value of $l$ for each sub-rod.

With this approximation, we find using Eq.~(\ref{eq:Ipkssum}) and (\ref{eq:Ipkstot}) that, at a given $l$, the ratio of the integrated intensity of the sharp peak from the $L$th sub-rod to the total intensity of all sharp peaks is

\begin{multline}
\frac{I_{\mathrm{p}}^{\mathrm{tot},L}}{I_{\mathrm{p}}^{\mathrm{tot}}} = \frac{\sin^2(\pi x)}{\pi^2x^2} e^{-x^2 \pi^2 \tilde{\sigma}_{\mathrm{tot}}^2 } \\
\bigg/\left(1 - \frac{4}{\sqrt{\pi}} \tilde{\sigma}_{\mathrm{tot}} \sin^2(\pi x) \right),
\label{eq:pkratio}
\end{multline}
where $x \equiv l - L$ is the distance along the sub-rod, in the $q_z$ direction, to the primary Bragg point. This ratio depends only on the total roughness. Thus, by fitting the observed ratios to this expression, shown in Fig.~\ref{fig:analysis}(d)-(f), we can extract $\tilde{\sigma}_{\mathrm{tot}}$. Even though the intensities of the sub-rods vary by many orders of magnitude, the ratios collapse onto a single curve, and the fit is excellent.

The easiest way to find $\sigma_{\mathrm{w}}$ would be to compare the integrated intensity in all of the sharp peaks to the integrated intensity in the broad peak. However, since the broad peak is often wide and weak, it is hard to accurately measure the integrated intensity. Thus, we focus on the maximum intensity. As shown in dashed box in Fig.~\ref{fig:schematic}(b), we define $I_{\mathrm{w}}^{\rm{max}}$ to be the sum of the intensities of the pixels in the column that contains the broad peak maximum. Using Eq.~(\ref{eq:Ipkstot}) and (\ref{eq:Iw}), we calculate the ratio

\begin{equation}
\frac{I_{\mathrm{w}}^{\rm{max}}}{I_{\mathrm{p}}^{\mathrm{tot}}} = \sin^2(\pi l) \frac{2 \tilde{\sigma}_{\mathrm{w}}^2}{\pi} \Delta \bigg/ \bigg( 1 - \frac{4}{\sqrt{\pi}} \tilde{\sigma}_{\mathrm{tot}} \sin^2(\pi l) \bigg),
\label{eq:wratio}
\end{equation}
where $\Delta$ is the width of a detector pixel in reciprocal space. This ratio depends on $\tilde{\sigma}_{\mathrm{w}}$ and $\tilde{\sigma}_{\mathrm{tot}}$. We plot the observed ratio as red squares in Fig.~\ref{fig:analysis}(a)-(c). Using Eq.~(\ref{eq:wratio}) and our best fit value for $\tilde{\sigma}_{\mathrm{tot}}$, we do a least squares fit to extract $\tilde{\sigma}_{\mathrm{w}}$.

\begin{table}[hp!]
\centering
\setlength{\tabcolsep}{8pt}
\begin{tabular}{p{3.5cm} c | c c c}
 & \multicolumn{1}{c}{} & \multicolumn{3}{c}{Sample} \\
 &  & A & B & C\\ \hline % & $M$ (unit cells)
\multirow{2}{3.5cm}{Best fit to diffraction pattern} & $\tilde{\sigma}_{\mathrm{w}}$ & 0.07 & 0.11 & 0.14 \\% \cline{2-5}
 & $\tilde{\sigma}_{\mathrm{s}}$ & 0 & 0 & 0.14 \\ [3pt] \hline
\multirow{2}{3.5cm}{AFM image, presuming $\xi_y = 100$ nm} & $\tilde{\sigma}_{\mathrm{w}}$ & 0.06 & 0.10 & 0.12 \\
 & $\tilde{\sigma}_{\mathrm{s}}$ & 0.01 & 0.02 & 0.17 \\
\end{tabular}
\caption{Comparison of roughness parameters extracted from truncation rod fitting and AFM images. Confidence intervals are less than $\pm 0.01$ for all parameters.}
\label{tab:roughness}
\end{table}

To complete the test of our model, we compare the roughness parameters extracted from truncation rod fitting to the roughness parameters found directly from the AFM images in Table~ \ref{tab:roughness}. To find the roughness parameters from the AFM images, we use a correlation length $\xi_y$ of 100~nm, inferred from the width of the truncation rods in $q_y$, and presume that the $\xi_x$ exceeds the width of the image. We calculate the average step position and the step edge jaggedness in horizontal 100 nm strips, and then average over all strips to find $\tilde{\sigma}_{\mathrm{w}}$ and $\tilde{\sigma}_{\mathrm{s}}$. As we show in Table \ref{tab:roughness}, the agreement between the two methods is excellent. However, $\tilde{\sigma}_{\mathrm{s}}$ from the AFM images systematically exceeds $\tilde{\sigma}_{\mathrm{s}}$ from the truncation rod fits, while the opposite is true for $\tilde{\sigma}_{\mathrm{w}}$. We hypothesize in the Supplemental Material\cite{SM} how correlations in the step edge jaggedness, which are not captured in our model, may be responsible for this discrepancy. Overall, the good agreement between the parameters extracted from the two methods indicates that our model successfully describes the terrace roughness of the three surfaces.

%%%%%
\section{Discussion}
\label{sec:Discussion}

\begin{figure}[htp!]
\includegraphics[width=3.375in]{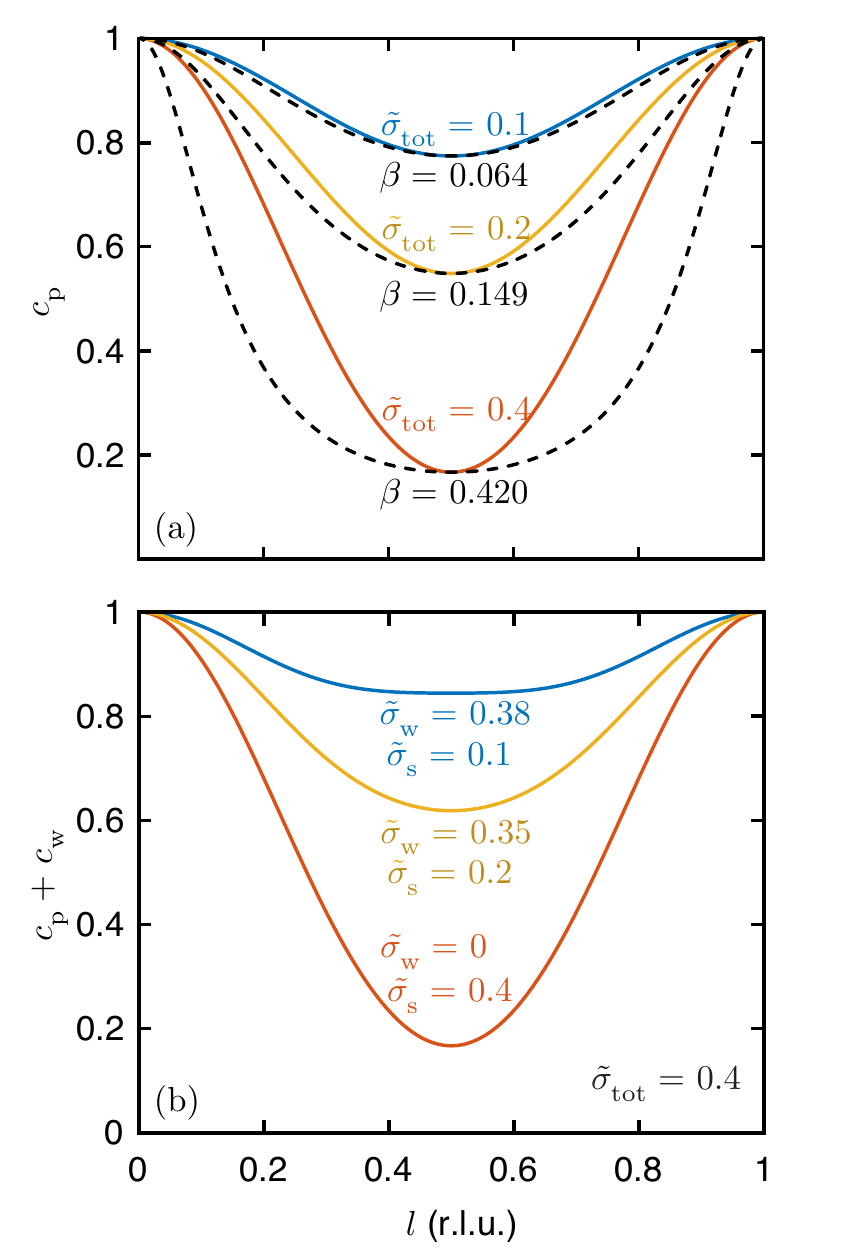}
\caption{Roughness factors. (a) $c_{\mathrm{p}}$, the ratio of the total intensity in the sharp peaks for a rough sample relative to an ideally miscut sample for various values of the total terrace roughness. The black dashed lines are $\beta$-roughness factors for comparison, calculated assuming $F_{\mathrm{s}} = F_{\mathrm{b}}$, chosen to match the $c_{\mathrm{p}}$ factor at $l = 0.5$. (b) The total intensity in the sharp peaks and the broad peak relative to the total intensity from an ideally miscut sample. The total roughness is the same for all curves, but it is split differently between terrace width variation and step edge jaggedness.}
\label{fig:corr}
\end{figure}

We have found that the truncation rods from miscut surfaces have three components: 1) a series of evenly spaced sharp peaks arising from the splitting of the truncation rod into sub-rods, 2) a single broad peak arising from terrace width variation, and 3) a diffuse background arising from step edge jaggedness. One of the most notable aspects of our model is the separation between solving the surface structure and evaluating the roughness. Indeed, we were able to characterize the terrace width and step edge roughness on three different samples without any knowledge of the surface or bulk structure factors, $F_{\mathrm{s}}$ and $F_{\mathrm{b}}$.

However, to do so, we had to examine in detail the intensities of the sub-rods and the broad peak. In a typical measurement, with lower resolution or smaller miscut, it might not be possible to resolve these peaks. As discussed in Sec.~\ref{sec:exp}, a typical analysis is likely to either 1) add integrated intensities from the sharp peaks and subtract $I_{\mathrm{w}}$ and $I_{\mathrm{s}}$ as background or 2) add integrated intensities from the sharp peaks and the broad peak, and subtract only $I_{\mathrm{s}}$ as background. In the first case, the rod intensity is $I_0$ times a roughness factor ($c_{\mathrm{p}}$) that depends only on $\sigma_{\mathrm{tot}}$ and $l$. This factor is shown in Fig.~\ref{fig:corr}(a). It is periodic in $l$, reaching unity at the Bragg points and a minimum at the anti-Bragg points. In the second case, the rod intensity is $I_0$ times a factor ($c_{\mathrm{p}} + c_{\mathrm{w}}$) that depends on $\sigma_{\mathrm{w}}$, $\sigma_{\mathrm{s}}$, and $l$. This factor is shown in Fig.~\ref{fig:corr}(b). It is also periodic in $l$, reaching a minimum at the anti-Bragg points, where the minimum value depends on the ratio of $\sigma_{\mathrm{w}}$ to $\sigma_{\mathrm{s}}$ in addition to $\sigma_{\mathrm{tot}}$. For two samples with the same $\sigma_{\mathrm{tot}}$, the sample with larger $\sigma_{\mathrm{w}}$ will have the shallower minimum.

The effect of terrace roughness is similar to other roughnesses because it decreases the intensity away from the Bragg peaks. As shown in Fig.~\ref{fig:corr}(a), the $l$ dependence is different than for $\beta$-roughness. The effect of terrace roughness is concentrated near the anti-Bragg point, whereas $\beta$-roughness results in a broader reduction in intensity. We note that multiple types of roughness may be present on a single sample. For example, it is possible for a surface to have $\beta$-roughness or partial occupancy across the entire surface in addition to having terrace roughness. In that case, $\beta$-roughness would only impact $I_0$, and terrace roughness would only impact $c_{\mathrm{p}}$ and $c_{\mathrm{w}}$.

%\section{Conclusion}
In conclusion, we have developed a new model for crystal truncation rods from miscut surfaces and applied it to a series of SrTiO$_3$ samples, where we characterized the terrace roughness without needing to solve the surface structure. Our model gives a simple multiplicative factor to account for this roughness in the crystal truncation rods. Our approach is broadly applicable to analyzing truncation rods from miscut samples and solving their surface structure.

% If in two-column mode, this environment will change to single-column format so that long equations can be displayed. 
% Use only when necessary.
%\begin{widetext}
%$$\mbox{put long equation here}$$
%\end{widetext}

% Tables may be be put in the text as floats.
% Here is an example of the general form of a table:
% Fill in the caption in the braces of the \caption{} command. Put the label
% that you will use with \ref{} command in the braces of the \label{} command.
% Insert the column specifiers (l, r, c, d, etc.) in the empty braces of the
% \begin{tabular}{} command.
%
% \begin{table}
% \caption{\label{} }
% \begin{tabular}{}
% \end{tabular}
% \end{table}

% If you have acknowledgments, this puts in the proper section head.
\vspace{10pt}
\begin{acknowledgments}
We thank Will Chueh and Yezhou Shi for providing a YSZ sample that inspired this work, and Kevin Stone for extensive discussion. Ron Marks and Bart Johnson provided indispensable assistance at the beamlines. This work was supported by Laboratory Directed Research and Development at the Stanford Synchrotron Radiation Lightsource, SLAC National Accelerator Laboratory, which is supported by the U.S. Department of Energy, Office of Science, Office of Basic Energy Sciences under Contract No. DE-AC02-76SF00515. AFM measurements were performed at the Stanford Nano Shared Facilities, supported by the National Science Foundation under award ECCS-1542152. T. P. was supported by the Department of Defense through a National Defense Science and Engineering Graduate Fellowship and by a William R. and Sara Hart Kimball Stanford Graduate Fellowship.
\end{acknowledgments}

% Create the reference section using BibTeX:
%\bibliography{paper4,paper4_supp}{}
%\bibliographystyle{plain}

%merlin.mbs apsrev4-1.bst 2010-07-25 4.21a (PWD, AO, DPC) hacked
%Control: key (0)
%Control: author (8) initials jnrlst
%Control: editor formatted (1) identically to author
%Control: production of article title (-1) disabled
%Control: page (0) single
%Control: year (1) truncated
%Control: production of eprint (0) enabled
%

% \bibliography{Mendeley.bib}

\begin{thebibliography}{31}%
\makeatletter
\providecommand \@ifxundefined [1]{%
 \@ifx{#1\undefined}
}%
\providecommand \@ifnum [1]{%
 \ifnum #1\expandafter \@firstoftwo
 \else \expandafter \@secondoftwo
 \fi
}%
\providecommand \@ifx [1]{%
 \ifx #1\expandafter \@firstoftwo
 \else \expandafter \@secondoftwo
 \fi
}%
\providecommand \natexlab [1]{#1}%
\providecommand \enquote  [1]{``#1''}%
\providecommand \bibnamefont  [1]{#1}%
\providecommand \bibfnamefont [1]{#1}%
\providecommand \citenamefont [1]{#1}%
\providecommand \href@noop [0]{\@secondoftwo}%
\providecommand \href [0]{\begingroup \@sanitize@url \@href}%
\providecommand \@href[1]{\@@startlink{#1}\@@href}%
\providecommand \@@href[1]{\endgroup#1\@@endlink}%
\providecommand \@sanitize@url [0]{\catcode `\\12\catcode `\$12\catcode
  `\&12\catcode `\#12\catcode `\^12\catcode `\_12\catcode `\%12\relax}%
\providecommand \@@startlink[1]{}%
\providecommand \@@endlink[0]{}%
\providecommand \url  [0]{\begingroup\@sanitize@url \@url }%
\providecommand \@url [1]{\endgroup\@href {#1}{\urlprefix }}%
\providecommand \urlprefix  [0]{URL }%
\providecommand \Eprint [0]{\href }%
\providecommand \doibase [0]{http://dx.doi.org/}%
\providecommand \selectlanguage [0]{\@gobble}%
\providecommand \bibinfo  [0]{\@secondoftwo}%
\providecommand \bibfield  [0]{\@secondoftwo}%
\providecommand \translation [1]{[#1]}%
\providecommand \BibitemOpen [0]{}%
\providecommand \bibitemStop [0]{}%
\providecommand \bibitemNoStop [0]{.\EOS\space}%
\providecommand \EOS [0]{\spacefactor3000\relax}%
\providecommand \BibitemShut  [1]{\csname bibitem#1\endcsname}%
\let\auto@bib@innerbib\@empty
%</preamble>
\bibitem [{\citenamefont {Robinson}\ and\ \citenamefont
  {Tweet}(1992)}]{Robinson1992}%
  \BibitemOpen
  \bibfield  {author} {\bibinfo {author} {\bibfnamefont {I.~K.}\ \bibnamefont
  {Robinson}}\ and\ \bibinfo {author} {\bibfnamefont {D.~J.}\ \bibnamefont
  {Tweet}},\ }\href {\doibase 10.1088/0034-4885/55/5/002} {\bibfield  {journal}
  {\bibinfo  {journal} {Rep. Prog. Phys.}\ }\textbf {\bibinfo {volume} {55}},\
  \bibinfo {pages} {599} (\bibinfo {year} {1992})}\BibitemShut {NoStop}%
\bibitem [{\citenamefont {Barbier}\ \emph {et~al.}(2000)\citenamefont
  {Barbier}, \citenamefont {Mocuta}, \citenamefont {Kuhlenbeck}, \citenamefont
  {Peters}, \citenamefont {Richter},\ and\ \citenamefont
  {Renaud}}]{Barbier2000}%
  \BibitemOpen
  \bibfield  {author} {\bibinfo {author} {\bibfnamefont {A.}~\bibnamefont
  {Barbier}}, \bibinfo {author} {\bibfnamefont {C.}~\bibnamefont {Mocuta}},
  \bibinfo {author} {\bibfnamefont {H.}~\bibnamefont {Kuhlenbeck}}, \bibinfo
  {author} {\bibfnamefont {K.~F.}\ \bibnamefont {Peters}}, \bibinfo {author}
  {\bibfnamefont {B.}~\bibnamefont {Richter}}, \ and\ \bibinfo {author}
  {\bibfnamefont {G.}~\bibnamefont {Renaud}},\ }\href {\doibase
  10.1103/PhysRevLett.84.2897} {\bibfield  {journal} {\bibinfo  {journal}
  {Physical Review Letters}\ }\textbf {\bibinfo {volume} {84}},\ \bibinfo
  {pages} {2897} (\bibinfo {year} {2000})}\BibitemShut {NoStop}%
\bibitem [{\citenamefont {Vonk}\ \emph {et~al.}(2005)\citenamefont {Vonk},
  \citenamefont {Konings}, \citenamefont {{Van Hummel}}, \citenamefont
  {Harkema},\ and\ \citenamefont {Graafsma}}]{Vonk2005}%
  \BibitemOpen
  \bibfield  {author} {\bibinfo {author} {\bibfnamefont {V.}~\bibnamefont
  {Vonk}}, \bibinfo {author} {\bibfnamefont {S.}~\bibnamefont {Konings}},
  \bibinfo {author} {\bibfnamefont {G.~J.}\ \bibnamefont {{Van Hummel}}},
  \bibinfo {author} {\bibfnamefont {S.}~\bibnamefont {Harkema}}, \ and\
  \bibinfo {author} {\bibfnamefont {H.}~\bibnamefont {Graafsma}},\ }\href
  {\doibase 10.1016/j.susc.2005.08.010} {\bibfield  {journal} {\bibinfo
  {journal} {Surface Science}\ }\textbf {\bibinfo {volume} {595}},\ \bibinfo
  {pages} {183} (\bibinfo {year} {2005})}\BibitemShut {NoStop}%
\bibitem [{\citenamefont {Feidenhans'l}\ \emph {et~al.}(1990)\citenamefont
  {Feidenhans'l}, \citenamefont {Grey}, \citenamefont {Johnson}, \citenamefont
  {Mochrie}, \citenamefont {Bohr},\ and\ \citenamefont
  {Nielsen}}]{Feidenhansl1990}%
  \BibitemOpen
  \bibfield  {author} {\bibinfo {author} {\bibfnamefont {R.}~\bibnamefont
  {Feidenhans'l}}, \bibinfo {author} {\bibfnamefont {F.}~\bibnamefont {Grey}},
  \bibinfo {author} {\bibfnamefont {R.~L.}\ \bibnamefont {Johnson}}, \bibinfo
  {author} {\bibfnamefont {S.~G.~J.}\ \bibnamefont {Mochrie}}, \bibinfo
  {author} {\bibfnamefont {J.}~\bibnamefont {Bohr}}, \ and\ \bibinfo {author}
  {\bibfnamefont {M.}~\bibnamefont {Nielsen}},\ }\href@noop {} {\bibfield
  {journal} {\bibinfo  {journal} {Physical Review B}\ }\textbf {\bibinfo
  {volume} {41}},\ \bibinfo {pages} {5420} (\bibinfo {year}
  {1990})}\BibitemShut {NoStop}%
\bibitem [{\citenamefont {Fenter}\ \emph {et~al.}(1994)\citenamefont {Fenter},
  \citenamefont {Eberhardt},\ and\ \citenamefont {Eisenberger}}]{Fenter1994}%
  \BibitemOpen
  \bibfield  {author} {\bibinfo {author} {\bibfnamefont {P.}~\bibnamefont
  {Fenter}}, \bibinfo {author} {\bibfnamefont {A.}~\bibnamefont {Eberhardt}}, \
  and\ \bibinfo {author} {\bibfnamefont {P.}~\bibnamefont {Eisenberger}},\
  }\href {\doibase 10.1126/science.266.5188.1216} {\bibfield  {journal}
  {\bibinfo  {journal} {Science}\ }\textbf {\bibinfo {volume} {266}},\ \bibinfo
  {pages} {1216} (\bibinfo {year} {1994})}\BibitemShut {NoStop}%
\bibitem [{\citenamefont {Willmott}\ \emph {et~al.}(2007)\citenamefont
  {Willmott}, \citenamefont {Pauli}, \citenamefont {Herger}, \citenamefont
  {Schlep{\"{u}}tz}, \citenamefont {Martoccia}, \citenamefont {Patterson},
  \citenamefont {Delley}, \citenamefont {Clarke}, \citenamefont {Kumah},
  \citenamefont {Cionca},\ and\ \citenamefont {Yacoby}}]{Willmott2007}%
  \BibitemOpen
  \bibfield  {author} {\bibinfo {author} {\bibfnamefont {P.~R.}\ \bibnamefont
  {Willmott}}, \bibinfo {author} {\bibfnamefont {S.~A.}\ \bibnamefont {Pauli}},
  \bibinfo {author} {\bibfnamefont {R.}~\bibnamefont {Herger}}, \bibinfo
  {author} {\bibfnamefont {C.~M.}\ \bibnamefont {Schlep{\"{u}}tz}}, \bibinfo
  {author} {\bibfnamefont {D.}~\bibnamefont {Martoccia}}, \bibinfo {author}
  {\bibfnamefont {B.~D.}\ \bibnamefont {Patterson}}, \bibinfo {author}
  {\bibfnamefont {B.}~\bibnamefont {Delley}}, \bibinfo {author} {\bibfnamefont
  {R.}~\bibnamefont {Clarke}}, \bibinfo {author} {\bibfnamefont
  {D.}~\bibnamefont {Kumah}}, \bibinfo {author} {\bibfnamefont
  {C.}~\bibnamefont {Cionca}}, \ and\ \bibinfo {author} {\bibfnamefont
  {Y.}~\bibnamefont {Yacoby}},\ }\href {\doibase 10.1103/PhysRevLett.99.155502}
  {\bibfield  {journal} {\bibinfo  {journal} {Physical Review Letters}\
  }\textbf {\bibinfo {volume} {99}},\ \bibinfo {pages} {155502} (\bibinfo
  {year} {2007})}\BibitemShut {NoStop}%
\bibitem [{\citenamefont {Yacoby}\ \emph {et~al.}(2013)\citenamefont {Yacoby},
  \citenamefont {Zhou}, \citenamefont {Pindak},\ and\ \citenamefont
  {Bo{\v{z}}ovi{\'{c}}}}]{Yacoby2013}%
  \BibitemOpen
  \bibfield  {author} {\bibinfo {author} {\bibfnamefont {Y.}~\bibnamefont
  {Yacoby}}, \bibinfo {author} {\bibfnamefont {H.}~\bibnamefont {Zhou}},
  \bibinfo {author} {\bibfnamefont {R.}~\bibnamefont {Pindak}}, \ and\ \bibinfo
  {author} {\bibfnamefont {I.}~\bibnamefont {Bo{\v{z}}ovi{\'{c}}}},\ }\href
  {\doibase 10.1103/PhysRevB.87.014108} {\bibfield  {journal} {\bibinfo
  {journal} {Physical Review B}\ }\textbf {\bibinfo {volume} {87}},\ \bibinfo
  {pages} {1} (\bibinfo {year} {2013})}\BibitemShut {NoStop}%
\bibitem [{\citenamefont {Toney}\ \emph {et~al.}(1994)\citenamefont {Toney},
  \citenamefont {Howard}, \citenamefont {Richer}, \citenamefont {Borges},
  \citenamefont {Gordon}, \citenamefont {Melroy}, \citenamefont {Wiesler},
  \citenamefont {Yee},\ and\ \citenamefont {Sorensen}}]{Toney1994}%
  \BibitemOpen
  \bibfield  {author} {\bibinfo {author} {\bibfnamefont {M.}~\bibnamefont
  {Toney}}, \bibinfo {author} {\bibfnamefont {J.}~\bibnamefont {Howard}},
  \bibinfo {author} {\bibfnamefont {J.}~\bibnamefont {Richer}}, \bibinfo
  {author} {\bibfnamefont {G.}~\bibnamefont {Borges}}, \bibinfo {author}
  {\bibfnamefont {J.}~\bibnamefont {Gordon}}, \bibinfo {author} {\bibfnamefont
  {O.}~\bibnamefont {Melroy}}, \bibinfo {author} {\bibfnamefont
  {D.}~\bibnamefont {Wiesler}}, \bibinfo {author} {\bibfnamefont
  {D.}~\bibnamefont {Yee}}, \ and\ \bibinfo {author} {\bibfnamefont
  {L.}~\bibnamefont {Sorensen}},\ }\href
  {http://www.nature.com/nature/journal/v368/n6470/abs/368444a0.html}
  {\bibfield  {journal} {\bibinfo  {journal} {Nature}\ }\textbf {\bibinfo
  {volume} {368}},\ \bibinfo {pages} {444} (\bibinfo {year}
  {1994})}\BibitemShut {NoStop}%
\bibitem [{\citenamefont {Liu}\ \emph {et~al.}(2016)\citenamefont {Liu},
  \citenamefont {Barbour}, \citenamefont {Komanicky},\ and\ \citenamefont
  {You}}]{Liu2016}%
  \BibitemOpen
  \bibfield  {author} {\bibinfo {author} {\bibfnamefont {Y.}~\bibnamefont
  {Liu}}, \bibinfo {author} {\bibfnamefont {A.}~\bibnamefont {Barbour}},
  \bibinfo {author} {\bibfnamefont {V.}~\bibnamefont {Komanicky}}, \ and\
  \bibinfo {author} {\bibfnamefont {H.}~\bibnamefont {You}},\ }\href {\doibase
  10.1021/acs.jpcc.6b00492} {\bibfield  {journal} {\bibinfo  {journal} {Journal
  of Physical Chemistry C}\ }\textbf {\bibinfo {volume} {120}},\ \bibinfo
  {pages} {16174} (\bibinfo {year} {2016})}\BibitemShut {NoStop}%
\bibitem [{\citenamefont {Eng}\ \emph {et~al.}(2000)\citenamefont {Eng},
  \citenamefont {Trainor}, \citenamefont {Brown}, \citenamefont {Waychunas},
  \citenamefont {Newville}, \citenamefont {Sutton},\ and\ \citenamefont
  {Rivers}}]{Eng2000}%
  \BibitemOpen
  \bibfield  {author} {\bibinfo {author} {\bibfnamefont {P.~J.}\ \bibnamefont
  {Eng}}, \bibinfo {author} {\bibfnamefont {T.~P.}\ \bibnamefont {Trainor}},
  \bibinfo {author} {\bibfnamefont {G.~E.}\ \bibnamefont {Brown}}, \bibinfo
  {author} {\bibfnamefont {G.~A.}\ \bibnamefont {Waychunas}}, \bibinfo {author}
  {\bibfnamefont {M.}~\bibnamefont {Newville}}, \bibinfo {author}
  {\bibfnamefont {S.~R.}\ \bibnamefont {Sutton}}, \ and\ \bibinfo {author}
  {\bibfnamefont {M.~L.}\ \bibnamefont {Rivers}},\ }\href {\doibase Doi
  10.1126/Science.288.5468.1029} {\bibfield  {journal} {\bibinfo  {journal}
  {Science}\ }\textbf {\bibinfo {volume} {288}},\ \bibinfo {pages} {1029}
  (\bibinfo {year} {2000})}\BibitemShut {NoStop}%
\bibitem [{\citenamefont {Andrews}\ and\ \citenamefont
  {Cowley}(1985)}]{Andrews1985}%
  \BibitemOpen
  \bibfield  {author} {\bibinfo {author} {\bibfnamefont {S.~R.}\ \bibnamefont
  {Andrews}}\ and\ \bibinfo {author} {\bibfnamefont {R.~A.}\ \bibnamefont
  {Cowley}},\ }\href {\doibase 10.1088/0022-3719/18/35/008} {\bibfield
  {journal} {\bibinfo  {journal} {Journal of Physics C: Solid State Physics}\
  }\textbf {\bibinfo {volume} {18}},\ \bibinfo {pages} {6427} (\bibinfo {year}
  {1985})}\BibitemShut {NoStop}%
\bibitem [{\citenamefont {Robinson}(1986)}]{Robinson1986}%
  \BibitemOpen
  \bibfield  {author} {\bibinfo {author} {\bibfnamefont {I.~K.}\ \bibnamefont
  {Robinson}},\ }\href {\doibase 10.1103/PhysRevB.33.3830} {\bibfield
  {journal} {\bibinfo  {journal} {Physical Review B}\ }\textbf {\bibinfo
  {volume} {33}},\ \bibinfo {pages} {3830} (\bibinfo {year}
  {1986})}\BibitemShut {NoStop}%
\bibitem [{\citenamefont {Sinha}\ \emph {et~al.}(1988)\citenamefont {Sinha},
  \citenamefont {Sirota}, \citenamefont {Garoff},\ and\ \citenamefont
  {Stanley}}]{Sinha1988}%
  \BibitemOpen
  \bibfield  {author} {\bibinfo {author} {\bibfnamefont {S.~K.}\ \bibnamefont
  {Sinha}}, \bibinfo {author} {\bibfnamefont {E.~B.}\ \bibnamefont {Sirota}},
  \bibinfo {author} {\bibfnamefont {S.}~\bibnamefont {Garoff}}, \ and\ \bibinfo
  {author} {\bibfnamefont {H.~B.}\ \bibnamefont {Stanley}},\ }\href {\doibase
  10.1103/PhysRevB.38.2297} {\bibfield  {journal} {\bibinfo  {journal}
  {Physical Review B}\ }\textbf {\bibinfo {volume} {38}},\ \bibinfo {pages}
  {2297} (\bibinfo {year} {1988})}\BibitemShut {NoStop}%
\bibitem [{\citenamefont {Shi}\ \emph {et~al.}(2016)\citenamefont {Shi},
  \citenamefont {Stone}, \citenamefont {Guan}, \citenamefont {Monti},
  \citenamefont {Cao}, \citenamefont {{El Gabaly}}, \citenamefont {Chueh},\
  and\ \citenamefont {Toney}}]{Shi2016}%
  \BibitemOpen
  \bibfield  {author} {\bibinfo {author} {\bibfnamefont {Y.}~\bibnamefont
  {Shi}}, \bibinfo {author} {\bibfnamefont {K.~H.}\ \bibnamefont {Stone}},
  \bibinfo {author} {\bibfnamefont {Z.}~\bibnamefont {Guan}}, \bibinfo {author}
  {\bibfnamefont {M.}~\bibnamefont {Monti}}, \bibinfo {author} {\bibfnamefont
  {C.}~\bibnamefont {Cao}}, \bibinfo {author} {\bibfnamefont {F.}~\bibnamefont
  {{El Gabaly}}}, \bibinfo {author} {\bibfnamefont {W.~C.}\ \bibnamefont
  {Chueh}}, \ and\ \bibinfo {author} {\bibfnamefont {M.~F.}\ \bibnamefont
  {Toney}},\ }\href {\doibase 10.1103/PhysRevB.94.205420} {\bibfield  {journal}
  {\bibinfo  {journal} {Physical Review B}\ }\textbf {\bibinfo {volume} {94}},\
  \bibinfo {pages} {1} (\bibinfo {year} {2016})}\BibitemShut {NoStop}%
\bibitem [{\citenamefont {Dale}\ \emph {et~al.}(2006)\citenamefont {Dale},
  \citenamefont {Fleet}, \citenamefont {Suzuki},\ and\ \citenamefont
  {Brock}}]{Dale2006}%
  \BibitemOpen
  \bibfield  {author} {\bibinfo {author} {\bibfnamefont {D.}~\bibnamefont
  {Dale}}, \bibinfo {author} {\bibfnamefont {A.}~\bibnamefont {Fleet}},
  \bibinfo {author} {\bibfnamefont {Y.}~\bibnamefont {Suzuki}}, \ and\ \bibinfo
  {author} {\bibfnamefont {J.~D.}\ \bibnamefont {Brock}},\ }\href {\doibase
  10.1103/PhysRevB.74.085419} {\bibfield  {journal} {\bibinfo  {journal}
  {Physical Review B}\ }\textbf {\bibinfo {volume} {74}},\ \bibinfo {pages} {1}
  (\bibinfo {year} {2006})}\BibitemShut {NoStop}%
\bibitem [{\citenamefont {Schlep{\"{u}}tz}\ \emph {et~al.}(2005)\citenamefont
  {Schlep{\"{u}}tz}, \citenamefont {Herger}, \citenamefont {Willmott},
  \citenamefont {Patterson}, \citenamefont {Bunk}, \citenamefont
  {Br{\"{o}}nnimann}, \citenamefont {Henrich}, \citenamefont {H{\"{u}}lsen},\
  and\ \citenamefont {Eikenberry}}]{Schleputz2005}%
  \BibitemOpen
  \bibfield  {author} {\bibinfo {author} {\bibfnamefont {C.~M.}\ \bibnamefont
  {Schlep{\"{u}}tz}}, \bibinfo {author} {\bibfnamefont {R.}~\bibnamefont
  {Herger}}, \bibinfo {author} {\bibfnamefont {P.~R.}\ \bibnamefont
  {Willmott}}, \bibinfo {author} {\bibfnamefont {B.~D.}\ \bibnamefont
  {Patterson}}, \bibinfo {author} {\bibfnamefont {O.}~\bibnamefont {Bunk}},
  \bibinfo {author} {\bibfnamefont {C.}~\bibnamefont {Br{\"{o}}nnimann}},
  \bibinfo {author} {\bibfnamefont {B.}~\bibnamefont {Henrich}}, \bibinfo
  {author} {\bibfnamefont {G.}~\bibnamefont {H{\"{u}}lsen}}, \ and\ \bibinfo
  {author} {\bibfnamefont {E.~F.}\ \bibnamefont {Eikenberry}},\ }\href
  {\doibase 10.1107/S0108767305014790} {\bibfield  {journal} {\bibinfo
  {journal} {Acta Crystallographica Section A: Foundations of Crystallography}\
  }\textbf {\bibinfo {volume} {61}},\ \bibinfo {pages} {418} (\bibinfo {year}
  {2005})}\BibitemShut {NoStop}%
\bibitem [{\citenamefont {Lu}\ and\ \citenamefont {Lagally}(1982)}]{Lu1982}%
  \BibitemOpen
  \bibfield  {author} {\bibinfo {author} {\bibfnamefont {T.~M.}\ \bibnamefont
  {Lu}}\ and\ \bibinfo {author} {\bibfnamefont {M.~G.}\ \bibnamefont
  {Lagally}},\ }\href {\doibase 10.1017/CBO9781107415324.004} {\bibfield
  {journal} {\bibinfo  {journal} {Surface Science}\ }\textbf {\bibinfo {volume}
  {120}},\ \bibinfo {pages} {47 } (\bibinfo {year} {1982})}\BibitemShut
  {NoStop}%
\bibitem [{\citenamefont {Presicci}\ and\ \citenamefont
  {Lu}(1984)}]{Presicci1984}%
  \BibitemOpen
  \bibfield  {author} {\bibinfo {author} {\bibfnamefont {M.}~\bibnamefont
  {Presicci}}\ and\ \bibinfo {author} {\bibfnamefont {T.~M.}\ \bibnamefont
  {Lu}},\ }\href {\doibase 10.1016/0039-6028(84)90208-5} {\bibfield  {journal}
  {\bibinfo  {journal} {Surface Science}\ }\textbf {\bibinfo {volume} {141}},\
  \bibinfo {pages} {233} (\bibinfo {year} {1984})}\BibitemShut {NoStop}%
\bibitem [{\citenamefont {Pukite}\ \emph {et~al.}(1985)\citenamefont {Pukite},
  \citenamefont {Lent},\ and\ \citenamefont {Cohen}}]{Pukite1985}%
  \BibitemOpen
  \bibfield  {author} {\bibinfo {author} {\bibfnamefont {P.}~\bibnamefont
  {Pukite}}, \bibinfo {author} {\bibfnamefont {C.}~\bibnamefont {Lent}}, \ and\
  \bibinfo {author} {\bibfnamefont {P.}~\bibnamefont {Cohen}},\ }\href
  {\doibase 10.1016/0167-2584(85)90512-2} {\bibfield  {journal} {\bibinfo
  {journal} {Surface Science Letters}\ }\textbf {\bibinfo {volume} {161}},\
  \bibinfo {pages} {A534} (\bibinfo {year} {1985})}\BibitemShut {NoStop}%
\bibitem [{\citenamefont {Croset}\ and\ \citenamefont
  {de~Beauvais}(1998)}]{Croset1998}%
  \BibitemOpen
  \bibfield  {author} {\bibinfo {author} {\bibfnamefont {B.}~\bibnamefont
  {Croset}}\ and\ \bibinfo {author} {\bibfnamefont {C.}~\bibnamefont
  {de~Beauvais}},\ }\href {\doibase 10.1016/S0039-6028(98)00182-4} {\bibfield
  {journal} {\bibinfo  {journal} {Surface Science}\ }\textbf {\bibinfo {volume}
  {409}},\ \bibinfo {pages} {403} (\bibinfo {year} {1998})}\BibitemShut
  {NoStop}%
\bibitem [{\citenamefont {Wollschl{\"{a}}ger}\ and\ \citenamefont
  {Tegenkamp}(2007)}]{Wollschlager2007}%
  \BibitemOpen
  \bibfield  {author} {\bibinfo {author} {\bibfnamefont {J.}~\bibnamefont
  {Wollschl{\"{a}}ger}}\ and\ \bibinfo {author} {\bibfnamefont
  {C.}~\bibnamefont {Tegenkamp}},\ }\href {\doibase 10.1103/PhysRevB.75.245439}
  {\bibfield  {journal} {\bibinfo  {journal} {Physical Review B}\ }\textbf
  {\bibinfo {volume} {75}},\ \bibinfo {pages} {1} (\bibinfo {year}
  {2007})}\BibitemShut {NoStop}%
\bibitem [{\citenamefont {Held}\ and\ \citenamefont {Brock}(1995)}]{Held1995}%
  \BibitemOpen
  \bibfield  {author} {\bibinfo {author} {\bibfnamefont {G.~A.}\ \bibnamefont
  {Held}}\ and\ \bibinfo {author} {\bibfnamefont {J.~D.}\ \bibnamefont
  {Brock}},\ }\href {\doibase 10.1103/PhysRevB.51.7262} {\bibfield  {journal}
  {\bibinfo  {journal} {Physical Review B}\ }\textbf {\bibinfo {volume} {51}},\
  \bibinfo {pages} {7262} (\bibinfo {year} {1995})}\BibitemShut {NoStop}%
\bibitem [{\citenamefont {Held}\ \emph {et~al.}(1995)\citenamefont {Held},
  \citenamefont {Goodstein},\ and\ \citenamefont {Brock}}]{Held1995b}%
  \BibitemOpen
  \bibfield  {author} {\bibinfo {author} {\bibfnamefont {G.~A.}\ \bibnamefont
  {Held}}, \bibinfo {author} {\bibfnamefont {D.~M.}\ \bibnamefont {Goodstein}},
  \ and\ \bibinfo {author} {\bibfnamefont {J.~D.}\ \bibnamefont {Brock}},\
  }\href {\doibase 10.1103/PhysRevB.51.7269} {\bibfield  {journal} {\bibinfo
  {journal} {Physical Review B}\ }\textbf {\bibinfo {volume} {51}},\ \bibinfo
  {pages} {7269} (\bibinfo {year} {1995})}\BibitemShut {NoStop}%
\bibitem [{\citenamefont {Pflanz}\ \emph {et~al.}(1995)\citenamefont {Pflanz},
  \citenamefont {Meyerheim}, \citenamefont {Moritz}, \citenamefont {Robinson},
  \citenamefont {Hoernis},\ and\ \citenamefont {Conrad}}]{Pflanz1995}%
  \BibitemOpen
  \bibfield  {author} {\bibinfo {author} {\bibfnamefont {S.}~\bibnamefont
  {Pflanz}}, \bibinfo {author} {\bibfnamefont {H.~L.}\ \bibnamefont
  {Meyerheim}}, \bibinfo {author} {\bibfnamefont {W.}~\bibnamefont {Moritz}},
  \bibinfo {author} {\bibfnamefont {I.~K.}\ \bibnamefont {Robinson}}, \bibinfo
  {author} {\bibfnamefont {H.}~\bibnamefont {Hoernis}}, \ and\ \bibinfo
  {author} {\bibfnamefont {E.~H.}\ \bibnamefont {Conrad}},\ }\href {\doibase
  10.1103/PhysRevB.52.2914} {\bibfield  {journal} {\bibinfo  {journal}
  {Physical Review B}\ }\textbf {\bibinfo {volume} {52}},\ \bibinfo {pages}
  {2914} (\bibinfo {year} {1995})}\BibitemShut {NoStop}%
\bibitem [{\citenamefont {Munkholm}\ and\ \citenamefont
  {Brennan}(1999)}]{Munkholm1999}%
  \BibitemOpen
  \bibfield  {author} {\bibinfo {author} {\bibfnamefont {A.}~\bibnamefont
  {Munkholm}}\ and\ \bibinfo {author} {\bibfnamefont {S.}~\bibnamefont
  {Brennan}},\ }\href {\doibase 10.1107/S0021889898005159} {\bibfield
  {journal} {\bibinfo  {journal} {Journal of Applied Crystallography}\ }\textbf
  {\bibinfo {volume} {32}},\ \bibinfo {pages} {143} (\bibinfo {year}
  {1999})}\BibitemShut {NoStop}%
\bibitem [{\citenamefont {Als-Nielsen}\ and\ \citenamefont
  {McMorrow}(2011)}]{Nielsen2011}%
  \BibitemOpen
  \bibfield  {author} {\bibinfo {author} {\bibfnamefont {J.}~\bibnamefont
  {Als-Nielsen}}\ and\ \bibinfo {author} {\bibfnamefont {D.}~\bibnamefont
  {McMorrow}},\ }\href {\doibase 10.1002/9781119998365} {\emph {\bibinfo
  {title} {{Elements of modern X-ray physics}}}}\ (\bibinfo {year}
  {2011})\BibitemShut {NoStop}%
\bibitem [{SM()}]{SM}%
  \BibitemOpen
  \href@noop {} {\bibinfo  {journal} {See Supplemental Material for a detailed
  derivation of the scattered intensity, a series expansion for the roughness
  factors, a discussion of miscut rotation, the Lorentz and illuminated area
  corrections, and limitations due to correlations}\ }\BibitemShut {NoStop}%
\bibitem [{\citenamefont {Kawasaki}\ \emph {et~al.}(1993)\citenamefont
  {Kawasaki}, \citenamefont {Maeda}, \citenamefont {Tsuchiya},\ and\
  \citenamefont {Koinuma}}]{Kawasaki1993}%
  \BibitemOpen
\bibfield  {journal} {  }\bibfield  {author} {\bibinfo {author} {\bibfnamefont
  {M.}~\bibnamefont {Kawasaki}}, \bibinfo {author} {\bibfnamefont
  {T.}~\bibnamefont {Maeda}}, \bibinfo {author} {\bibfnamefont
  {R.}~\bibnamefont {Tsuchiya}}, \ and\ \bibinfo {author} {\bibfnamefont
  {H.}~\bibnamefont {Koinuma}},\ }\href {\doibase
  10.1126/science.266.5190.1540} {\bibfield  {journal} {\bibinfo  {journal}
  {Science}\ }\textbf {\bibinfo {volume} {266}},\ \bibinfo {pages} {1}
  (\bibinfo {year} {1993})}\BibitemShut {NoStop}%
\bibitem [{\citenamefont {Busing}\ and\ \citenamefont
  {Levy}(1967)}]{Busing1967}%
  \BibitemOpen
  \bibfield  {author} {\bibinfo {author} {\bibfnamefont {W.~R.}\ \bibnamefont
  {Busing}}\ and\ \bibinfo {author} {\bibfnamefont {H.~A.}\ \bibnamefont
  {Levy}},\ }\href {\doibase 10.1107/S0365110X67000970} {\bibfield  {journal}
  {\bibinfo  {journal} {Acta Crystallographica}\ }\textbf {\bibinfo {volume}
  {22}},\ \bibinfo {pages} {457} (\bibinfo {year} {1967})}\BibitemShut
  {NoStop}%
\bibitem [{\citenamefont {Vlieg}(1997)}]{Vlieg1997}%
  \BibitemOpen
  \bibfield  {author} {\bibinfo {author} {\bibfnamefont {E.}~\bibnamefont
  {Vlieg}},\ }\href {\doibase 10.1107/S0021889897002537} {\bibfield  {journal}
  {\bibinfo  {journal} {Journal of Applied Crystallography}\ }\textbf {\bibinfo
  {volume} {30}},\ \bibinfo {pages} {532} (\bibinfo {year} {1997})}\BibitemShut
  {NoStop}%
\bibitem [{\citenamefont {You}(1999)}]{You1999}%
  \BibitemOpen
  \bibfield  {author} {\bibinfo {author} {\bibfnamefont {H.}~\bibnamefont
  {You}},\ }\href {\doibase 10.1107/S0021889899001223} {\bibfield  {journal}
  {\bibinfo  {journal} {Journal of Applied Crystallography}\ }\textbf {\bibinfo
  {volume} {32}},\ \bibinfo {pages} {614} (\bibinfo {year} {1999})}\BibitemShut
  {NoStop}%
\end{thebibliography}

%merlin.mbs apsrev4-1.bst 2010-07-25 4.21a (PWD, AO, DPC) hacked
%Control: key (0)
%Control: author (8) initials jnrlst
%Control: editor formatted (1) identically to author
%Control: production of article title (-1) disabled
%Control: page (0) single
%Control: year (1) truncated
%Control: production of eprint (0) enabled

\end{document}